\documentclass[conference]{IEEEtran}
\usepackage{amsmath}
\usepackage{algorithmic}
\usepackage{algorithm}
\usepackage{graphicx}
\usepackage{epstopdf}
\usepackage{epsfig}

\ifCLASSINFOpdf

\else

\fi

\begin{document}

\title{DEARF: Delay and Energy Aware RAW Formation Scheme to Support Delay Sensitive M2M Traffic in IEEE 802.11ah Networks}
\author{
\IEEEauthorblockN{Navroz Firoz Charania, Mukesh Kumar Giluka, Bheemarjuna Reddy Tamma and Antony Franklin}
\IEEEauthorblockA{Indian Institute of Technology Hyderabad,\\Email: [cs14mtech11003, cs11p1002, tbr and antony.franklin]~@iith.ac.in}
}

\maketitle

\begin{abstract}
The IEEE 802.11ah amendment is designed to support upto 8K M2M devices over the Sub-GHz channel. To achieve this, it introduces new modifications to the PHY and MAC layers. A dynamic Restricted Access Window~(RAW) mechanism is introduced at the MAC layer. RAW splits the access for different devices into small chunks of time. Using the RAW mechanism, we propose a novel Delay and Energy Aware RAW Formation~(DEARF) scheme to support delay sensitive devices along with other delay tolerant devices. We exploit this to introduce four new RAWs. These new RAWs help split the access into contention free for delay sensitive devices and contention based for other devices. For scheduling resources between these devices, we give the DEARF resource allocation and evaluate the scheme with help of simulations. Devices suffer a higher delay but, at high loads, devices using the proposed scheme are able to save 30\% energy per packet and upto 330\% energy per device. Packet delivery ratio is 100\% at low loads and 90\% at high loads as compared to 95\% and 27\% given by the traditional access. Other delay tolerant devices save upto 16\% on energy per packet transmitted and 23\% on delay at high loads.
\end{abstract}

\IEEEpeerreviewmaketitle

\section{Introduction}
\label{intro}
Nowadays, Internet of Things~(IoT) is trending among technological communities because of its scope and capability to influence human life style. Machine-to-Machine communication~(M2M) deals with communication and networking section of IoT. Characteristics of M2M communication mainly include support of large number of low power, low cost, low rate and less frequent communicating devices. Presently, almost all RATs~(radio access technologies) have developed their standards to support M2M features within their limitations. For example, bluetooth low energy~(BLE), 6LoWPAN and Zigbee. In the similar way, IEEE 802.11 Wi-Fi standard has amended its PHY and MAC layer to accommodate M2M characteristics and the standard is named as IEEE 802.11ah~(.11ah)~\cite{standard}. The PHY layer properties of .11ah are summarized in Table~\ref{phy} while the MAC layer properties are discussed in Section~\ref{overview}. 

\begin{table}[htb!]
\vspace{-0.5cm}
\caption{802.11ah PHY parameters}
\begin{center}

\begin{tabular}{|p{3.5cm}| p{3.5cm}|}

\hline

Frequency band & Sub-1GHz  \\
\hline
Range & upto 1Km \\
\hline
Number of devices supported& upto 8000 \\
\hline
Data rate & 100Kb/s \\
\hline
Topology & Single Hop \\
\hline
Per channel bandwidth & 1, 2, 4, 8 or 16MHz \\
\hline
Modulation schemes & BPSK, QPSK, 16-QAM, 64-QAM, 256-QAM \\
\hline
\end{tabular}
\label{phy}
\end{center}
\end{table}

In order to support a large number of M2M devices, IEEE 802.11ah standard distributes devices into multiple groups and each group is restricted to access the medium for a limited amount of time without any interference of other groups. The grouping of devices can be done based on various parameters such as geographical location of devices, application specific, power requirement, etc. The time window for which a group can access the channel is called as Restricted Access Window~(RAW). Devices belonging to a group not having access permission in the current RAW, go for sleeping and wake up in the beginning of their corresponding RAW, thereby saving energy. Within a RAW, devices follow CSMA/CA as the medium access protocol. IEEE 802.11ah~(.11ah) MAC protocol can be termed as hybrid protocol where TDMA is followed in the form of RAWs and CSMA/CA is followed within each RAW. If number of devices are more, then a RAW can further be divided into multiple slots so that collisions will be less and more energy per node can be saved. 
Apart from RAW, various existing features of IEEE 802.11 MAC layer have been modified to encourage energy saving and to facilitate the access of large number of devices. Some of these IEEE 802.11ah features are Synchronization Frame, Target Wait Time~(TWT), Bidirectional TXOP, Null Data Packet~(NDP), Short MAC Frame, Increased Sleep Time and hierarchical Traffic Indication Map~(TIM)~\cite{wi-fi-approach}. 

There are numerous IoT applications which .11ah can support. Some of these include smart grid, health care monitoring, environment and agriculture monitoring, industrial automation, etc. These applications may have delay sensitive data. However, the existing .11ah standard does not have support for such applications. In this paper, we propose a novel Delay and Energy Aware RAW Formation scheme~(DEARF) which ensures to meet delay constraints of delay sensitive applications as well as satisfactorily maintains the throughput of delay tolerant applications. The scheme concieves two types of access: contention free access to be dedicated to devices having delay sensitive data and contention based access for delay tolerant devices. The contention free access helps not only to meet the deadline of delay sensitive packets but also save energy of these devices, which would have otherwise, been wasted in case of contention based access due to channel sensing and retransmissions.


%
\section{Related Work}
\label{related}
\vspace{-0.2cm}
In this section we have discussed all recent works related to .11ah. In~\cite{sun} and~\cite{feasibility}, the authors focus on discussing the PHY layer aspects of .11ah. Authors of~\cite{sun}, present a technical overview of PHY and MAC layers of .11ah. They discuss MAC enhancements made in Wi-Fi to support M2M communications and evaluates the performance of .11ah in terms of the transmission range of the devices and overall throughput. The authors of~\cite{feasibility} discuss path loss models, shadowing and fast fading effect, achievable data rate in terms of bit error rate, bit rate comparison for shadowing and fast fading scenario along packet size design with PHY constraints. 

In~\cite{wi-fi-approach},~\cite{park},~\cite{performance-enhancements} and~\cite{performance-evaluation}, the authors discuss the MAC layer performance of IEEE 802.11ah. The authors of~\cite{wi-fi-approach}, discuss MAC contributions viz., grouping of devices and beacon signals. Performance evaluation of IEEE 802.11ah is done based on four M2M scenarios viz., agriculture monitoring, smart metering, industrial automation and animal monitoring. The authors of~\cite{park} discuss the hidden terminal problem and probe delay problem in IEEE 802.11ah. They discuss the hidden terminal problem, probe delay problem and their solution by SMA and usage of $Synch$ frame is also stated. Authors evaluate the performance gains of these solutions. The authors of~\cite{performance-enhancements} evaluate the performance of individual and coinciding IEEE 802.11ah networks through simulations. They compare the performance of DCF with basic access and DCF with RTS/CTS schemes in .11ah. Basic access turns out to be more efficient. Later, they evaluate IEEE 802.11ah networks having multiple Access Points~(APs), overlapping with each other, based on metrics such as throughput and energy consumption. In~\cite{performance-evaluation}, an analytical model for saturation throughput and energy consumption for IEEE 802.11ah systems is developed. The performance is evaluated in terms of throughput, delay and energy consumption. The authors of~\cite{zigbee} compare IEEE 802.11ah with IEEE 802.15.4. IEEE 802.11ah standard outperforms the latter in throughput in both idle and erroneous medium conditions. They compare for energy consumption by varying the system from having light load to dense load. Devices in .11ah consume lesser energy in light load but as load increases IEEE 802.15.4 starts to outperform.

In~\cite{cas},~\cite{holding} and~\cite{alarm}, the authors extend IEEE 802.11ah by proposing enhancements at the MAC layer. In~\cite{cas}, an optimization problem is solved which is to optimize the number of channel access slots, their length and their allocation to the stations for both uplink and downlink. It ensures a high reliability in terms of packet delivery along with maximizing the time that devices remain in the sleep state. In~\cite{holding}, authors discuss the different holding schemes for non-cross slot boundary transmissions and propose two new holding schemes. In~\cite{alarm}, the authors propose a hybrid scheme optimized for serving event driven devices. The devices are given a few contention free resources at the start and as the devices reporting increases, the number of resources are increased. This happens in the same beacon interval dynamically. We see, that the approach given in the paper helps in serving the triggered events sooner than the basic approach, with a slow down in performance for non-event driven devices.

Both the .11ah standard~\cite{standard} and currently available literature do not discuss much on the issues of providing reliable and in time packet delivery. Apart from the inherent energy savings provided by introducing RAW access mechanism and IEEE 802.11e amendment, there is not much research available on further reducing energy. We address the issue of providing reliable delivery to delay sensitive applications, along with energy savings. .11ah implements a combination of both TDMA and CSMA/CA for medium access. Exploiting this, we provide two distinct types of access viz., contention free and contention based. Contention free access is used for serving delay sensitive devices and contention based for other regular devices. As shown in~\cite{collisions}, the AP can distinguish between collisions from propagation errors. With this information, we later give an algorithm to manage resources between the two access periods stated above.

\section{Overview of IEEE 802.11ah MAC Layer}
\label{overview}
The `ah' amendment to IEEE 802.11 standard introduces changes to the MAC layer to support M2M communications. In .11ah, medium access is scheduled in terms of one beacon interval at a time. The interval is divided into several Restricted Access Windows~(RAWs) and a group~(or groups) of devices is assigned a RAW for access. The RAWs can be divided into slots to further reduce contention. Actually, a beacon interval is the time between two consecutive Delivery Transmission Indication Map~(DTIM) beacons. DTIM beacons are sent by AP to devices at fixed interval~(at the beginning of DTIM interval). Each device in a .11ah network, having data to transmit or receive, has to listen to this beacon. DTIM beacon carries information about time of arrival of the RAWs and grouping information of devices. Traffic Indication Map~(TIM) beacon is sent by AP in the beginning of each RAW which is received by all devices of a group~(or groups) to which the corresponding RAW is assigned. TIM beacons provide information about mapping devices to slots within the RAW. They also indicate the presence of a downlink data packet to a device and indicate the slot in which a device may make a PS-Poll to retrieve the packet.
\begin{figure}[htb!]
 \epsfig{width=9cm,figure=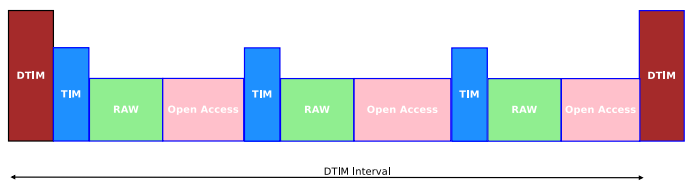}
 \caption{Structure of a DTIM Interval}
 \label{DTIM}
 \end{figure}
 
 \begin{figure}[htb!]
 \centering
 \epsfig{width=6cm,height=3cm,figure=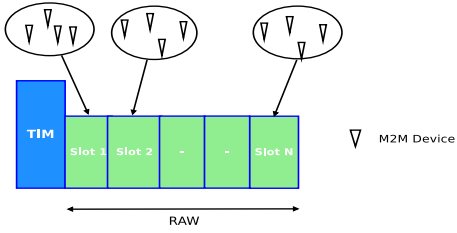}
 \caption{RAW Structure}
 \label{raw}
 \vspace{-0.5cm}
 \end{figure}
Figure~\ref{DTIM} shows a typical DTIM interval which starts with a DTIM beacon and TIM beacons arriving before the start of each RAW. The interval may also include open access for usage by legacy devices. Figure~\ref{raw} shows the structure of a RAW which consists of $N$ slots and has different set of devices mapped to different slots. Devices cannot access the medium before the
beginning of their designated slots. The formula used to decide the slot in which a device can access the medium is given by the following equation:
\begin{equation}
X_{slot}=(AID_{X}+N_{offset})mod(N)
\end{equation}
where $X_{slot}$ is the slot number of the device $X$ whose AID~(Association Identifier) is $AID_X$. $N_{offset}$ is an offset used to improve fairness and is calculated by the AP. 

Devices having downlink or uplink data first wake up to receive the DTIM beacon and then wake up to receive its designated TIM beacon. Through the TIM beacons, slot numbers of the devices are communicated and devices access the medium in the given slot by following DCF. If a device is not able to send its data in the current slot then the it will wait till the arrival of its designated slot in the next DTIM interval. Size of a RAW and its slot duration can be varied in every DTIM interval.

\section{Proposed Work}
\label{proposed}
As discussed in Section~\ref{overview}, devices perform CSMA/CA to send uplink data in their designated RAW slots. A slot will usually contain multiple devices contending to send data. In case of a high load in the network, it is highly probable that a large number of collisions will take place in a slot, which may result in unprecedented increase in end-to-end packet delay. This creates an unfavorable condition for mission critical or delay sensitive applications to be supported by .11ah. Apart from this, due to excess collisions, multiple transmissions of the same packet will take place. This will cause unnecessary power consumption by M2M devices. In this section, we propose a novel Delay and Energy Aware RAW Formation~(DEARF) scheme to meet the time constraints of delay sensitive devices as well as to minimize energy consumptions for both delay sensitive and delay tolerant devices. In this scheme, delay intolerant devices are given contention free slots to avoid delay as well as to save energy~(by avoiding retransmission of packets). While, delay tolerant devices are allowed to continue their access, by following basic .11ah access mechanism. Here, contention for delay tolerant devices is also reduced as delay intolerant devices are removed from contending for the medium. 

In this paper, we assume two types of M2M devices. Firstly the Delay Sensitive Machine type Devices~(DSMDs) having delay sensitive uplink data which must be delivered to the AP within a given deadline~(in order of DTIM periods). And then, non-Delay Sensitive Machine type Devices~(non-DSMDs) which have no such delay constraints. As discussed earlier, an AP allocates radio resources in time fractions of a DTIM interval and a DTIM interval is occupied by multiple RAWs having multiple slots each. In our work, we define four type of RAWs, each of them has different size and functionalities. These RAWs are explained as follows:

\textbf{Contention Indication~(CI) RAW:} This RAW is used to indicate to the AP that DSMDs have data to send. The RAW is divided into multiple slots, each slot is assigned to a group of DSMDs. Only DSMDs of a particular group are allowed to access in the corresponding CI RAW slot. If a DSMD has data to send, it transmits a beacon signal to AP in the slot belonging to that DSMD's group. Similarly, other DSMDs of that group having data will send beacons in the same CI slot resulting in a collision. On realizing this signal or collision in a slot of the CI RAW, the AP will know that DSMDs of the corresponding group have delay sensitive data. The AP, then allocates contention free slots, equal to the number of DSMDs in the corresponding group, in the DII RAW~(discussed next). The size of a CI slot is the sum of transmission and propagation times of a simple beacon packet. The size of a CI RAW is the product of CI slot size with number of DSMD groups.

\textbf{Delay Information Indication~(DII) RAW:} On noticing contention in CI RAW, the AP creates DII RAW in the same DTIM. A DII RAW contains only contention free slots. One CI slot is mapped to one group of DSMDs. So, on noticing contention in any of the slots of the CI RAW, the AP will allocate DII slots to each DSMD of the groups corresponding to these CI slots. Therefore, the number of contention free slots in DII RAW for a given DTIM interval will be the total number of DSMDs from all groups in which contention was noticed in the preceding CI RAW. In the contention free DII slot, the DSMDs send a small control packet~(10 Bytes) to the AP carrying the information of packet delay requirements and time of arrival. The existence of DII RAW in a DTIM depends on the presence of contention in CI RAW. In case of contention in CI RAW, only single DII RAW will exist in a DTIM otherwise, no DII RAW will exist for the current DTIM. 

\textbf{DSMDs Resource Allocation~(DRA) RAW:} After receiving delay information from DSMDs in DII RAW of current DTIM, the AP is able to estimate the exact number of DSMDs wanting to send their data. The AP assigns contention free slots to these DSMDs in the DRA RAW of following DTIMs. The number of DRA RAWs can be more than one depending on number of DSMDs having data to send and maximum limit on the size of a RAW. A DRA RAW may not exist if in previous DTIMs DSMDs have not made any request for transmission.

\textbf{Non-DSMDs Resource Allocation~(NRA) RAW:} This RAW is used to allocate resources to non-DSMDs. It provides contention based slots in which non-DSMDs contend for the channel by following the usual Distributed Co-ordination Function~(DCF). The number of NRA RAWs in a DTIM can be more than one. A DTIM will contain at least one NRA RAW for serving non-DSMDs.

\begin{figure}[htb!]
 \epsfig{width=9cm,figure=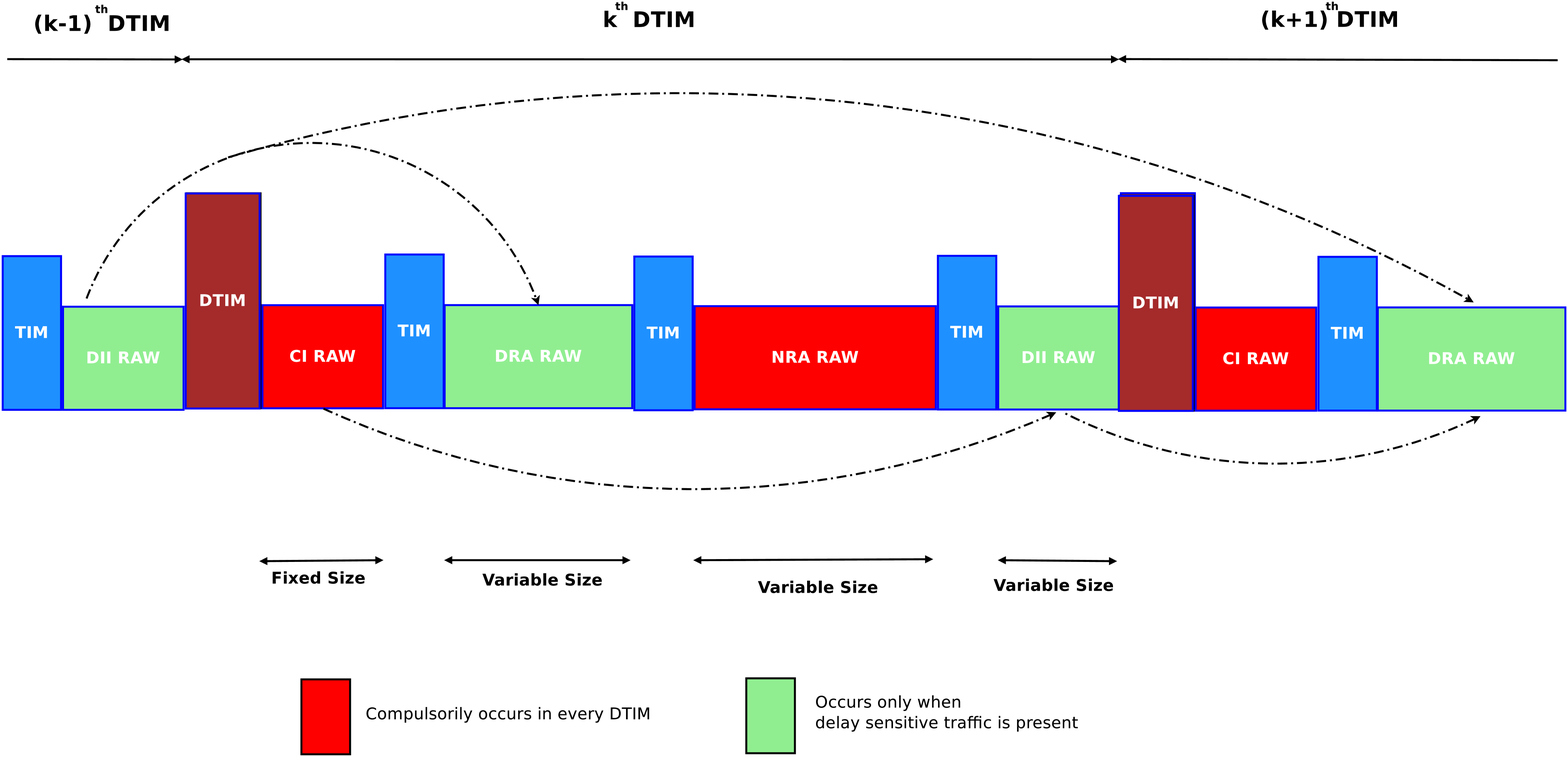}
 \caption{Proposed DTIM Structure}
 \label{proposed_dtim}
 
 \end{figure}
 
\begin{figure}[htb!]
 \epsfig{width=8cm,figure=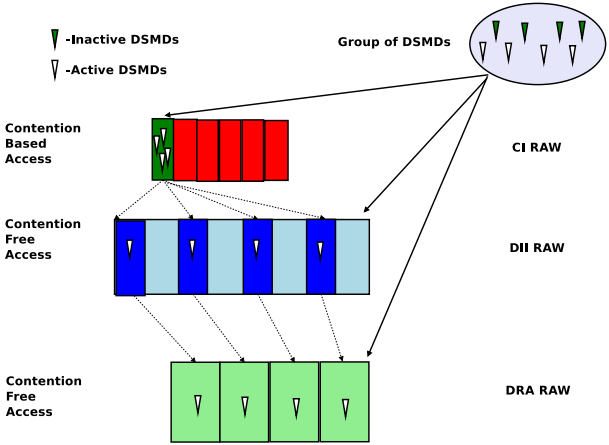}
 \caption{Proposed DEARF Scheme}
 \label{DEARF}
 \vspace{-0.5cm}
 \end{figure}



A DTIM interval contains four types of RAWs, in which first three are used to serve DSMDs while the fourth is used to serve non-DSMDs. In the beginning of each DTIM, all DSMDs and non-DSMDs receive a DTIM beacon. In a DTIM interval, two categories of DSMDs exist in which first are those which access the medium in CI and DII RAWs and second are those which access in DRA RAW. DSMDs which belong to the first category are termed as $contending$ DSMDs whereas the second category of DSMDs are termed as $transmitting$ DSMDs. $Contending$ DSMDs of previous DTIM become $transmitting$ DSMDs of current DTIM. Once the status of a DSMD changes from $contending$ to $transmitting$, it will remain same until the DSMD is not served or deadline is crossed.

The first RAW in the proposed DTIM structure is the CI RAW, which always accessed by $contending$ DSMDs. Hence, to indicate the arrival of a CI RAW, no TIM beacon is required. $Contending$ DSMDs receive the time of arrival of TIM of DII RAW through a special beacon right before the start of NRA RAW. Slotting of DII RAW is done after observing the contention in the CI RAW. Hence, the size of DII RAW will be variable depending on number of slots of CI RAW~(in the current DTIM) where contention was observed. $Transmitting$ DSMDs receive the time of arrival of TIM of DRA RAW. All non-DSMDs receive time of arrival of TIM of NRA RAW through the DTIM beacon. Resources to be allocated to DRA and NRA RAWs of a DTIM are decided in the beginning of DTIM. The decision is conveyed to the devices through DTIM beacons. Figure~\ref{proposed_dtim} shows the proposed DTIM interval. The arrows indicate the flow in which DSMDs access the medium.

Figure~\ref{DEARF} illustrates the proposed DEARF scheme. Here, a group of eight DSMDs is present in which four DSMDs have data to send, therefore, these perform contention based medium access in CI RAW of current DTIM. After noticing contention in one of the slots of CI RAW, the AP creates a DII RAW in the same DTIM containing eight contention free slots. Here, AP assumes that all DSMDs of the group have data to send. Now, DSMDs having delay sensitive data~(in this case four DSMDs), submit delay constraint information of their data to the AP in the dedicated contention free slot of DII RAW in the same DTIM. In the DII RAW of the figure, in blue colored contention free slots, DSMDs send delay constraint information of their packet whereas sky color slots go empty. By observing occupied contention free slots in the DII RAW of the current DTIM, AP calculates the exact number of $transmitting$ DSMDs with respective delay information. Since four blue colored slots are there in DII RAW, hence, four $transmitting$ DSMDs are there for whom AP will allocate contention free slots in DRA RAW of subsequent DTIMs for uplink data transmissions as shown in Figure~\ref{proposed_dtim}. 
In order to distribute resources among non-DSMDs and $transmitting$ DSMDs, we have developed an algorithm which allocates resources based on delay constraints of $transmitting$ DSMDs and at the same time ensures throughput of non-DSMDs. $Transmitting$ DSMDs are allocated dedicated contention free slots in DRA RAW and non-DSMDs are allocated contention based slots in NRA RAW. Parameters $C$, $T$, $\lambda$, $N_{collision}$ are passed as input to Algorithm~\ref{algorithm1}. Here, $C$ is set of classes where $C_{i}$ is a class of $transmitting$ DSMDs whose deadline is going to end at $i^{th}$ DTIM from the current DTIM. For example, DSMDs of class $C_{0}$ must be served in the current DTIM. $T$ is the minimum resource~(in terms of a fraction of DTIM interval) which must be assigned to non-DSMDs~(NRA RAW). $N_{collision}$ is the total number of collisions occurred in NRA RAWs from the last DTIM. $\lambda$ is a threshold used for deciding on allocating extra resources to NRA. If value of $N_{collisions}$ is greater than $\lambda$ then the algorithm allocates more resources to non-DSMDs in current DTIM. The output is the updated classes $C$ and allocation of resources to DSMDs.

\renewcommand{\algorithmicrequire}{\textbf{Input:}}
\renewcommand{\algorithmicensure}{\textbf{Output:}}
\begin{algorithm}

\caption{DEARF Resource Allocation Algorithm}
\label{algorithm1}
\begin{algorithmic}[1]
\REQUIRE $C$, $T$, $\lambda$, $N_{collision}$
\ENSURE Allocation of resources to devices, updated $C$
\STATE $T_{Avail}\leftarrow$ Available DTIM time for DRA and NRA RAWs
 \STATE Allocate resources to \textit{transmitting} DSMDs of $C_0$
 \STATE Allocate $T$ amount of resources to non-DSMDs
 \IF {$N_{collision}\geq \lambda$ and $T_{Avail} \neq 0$ }
  \STATE Allocate remaining resources to non-DSMDs by increasing NRA RAWs
  \ELSE
  \STATE Allocate remaining resources to \textit{transmitting} DSMDs of $C_1$, $C_2$, \dots
  
\ENDIF
\STATE Update $C$
\end{algorithmic}

\end{algorithm}

\section{Simulation Results and Analysis}
\label{results}
In this section, we have compared and analyzed the performance of proposed DEARF scheme with respect to traditional .11ah medium access scheme, termed as Basic  scheme, through MATLAB simulations. Metrics used to evaluate the performance include packet delay, energy consumption of devices and packet delivery ratio. Table~\ref{basic_parameters} shows .11ah PHY and MAC layer simulation parameters common to both schemes and Table~\ref{DEARF_parameters} shows additional parameters for DEARF scheme.

We consider three data arrival scenarios. In the scenario ``Arrival within $X$ DTIM $Scheme$'', $X$ denotes the no. of DTIMs across which DSMDs have data arrival, $Scheme$ will be either Basic or DEARF. Values of $X$ are taken as 1, 3, and 5. Lesser the value of $X$, more bursty the traffic in the network.
\begin{table}[htb!]
\vspace{-0.5cm}
 \caption{802.11ah MAC and PHY parameters common to both schemes}
 \begin{center}
 \vspace{-0.3cm}
\begin{tabular}{|c|p{3cm}|}
\hline
 Parameters & Values \\
 \hline \hline
 $CW_{min}$, $CW_{max}$, Retry Limit & $15$, $1023$, $4$\\
 \hline
 Simulation time & 18s\\
 \hline
 SIFS, DIFS & $160\mu s$, $274\mu s$\\
 \hline
 PHY Rate & MCS0 650Kbps\\
 \hline
 Packet size & 100 Bytes\\
 \hline
 DTIM interval & 1.6s\\
 \hline
 Size of DTIM and TIM beacons & 102 Bytes and 62 Bytes\\
 \hline
 RAW size and slot size & 200ms, 19ms\\
 \hline
  Number of non-DSMDs per DTIM& 200\\
 \hline
 Number of DSMDs per DTIM & 200, 400, 600, 800 and 1000\\ \hline
 Energy consumption: Rx, Tx, Idle and Sleep & 145mW, 285mW, 70mW and 5mW~\cite{zigbee}\\ \hline
\end{tabular}
\label{basic_parameters}
\end{center}
\end{table}
\begin{table}
\vspace{-0.5cm}
 \caption{DEARF Scheme simulation parameters}
 \begin{center}
\begin{tabular}{|c|c|}
\hline
 Parameters & Values \\
 \hline \hline
 CI RAW size and slot sizes& 18ms and 180$\mu$s\\
 \hline
 DII RAW slot size & 240$\mu$s\\
 \hline
 DRA RAW slot size & 1684$\mu$s\\
 \hline
 NRA RAW slot size & 19ms\\
 \hline
 Size of CI beacon and DII information Packet & 10 Bytes each\\
 \hline
\end{tabular}
\label{DEARF_parameters}
\end{center}
\end{table}

\begin{figure}[htb!]
\begin{minipage}{9cm}
 \centering
 \epsfig{width=6cm,figure=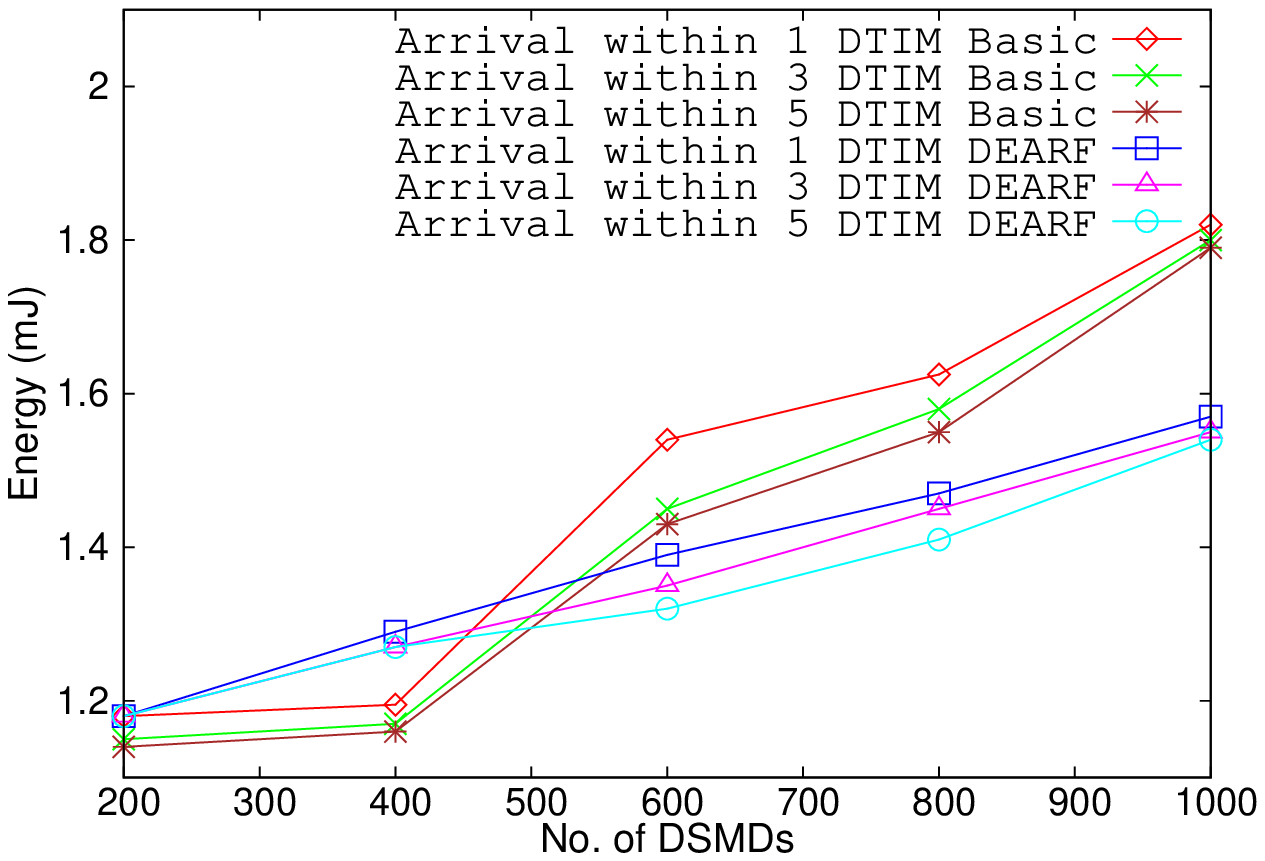}
 \caption{Average energy consumed per packet for DSMDs.}
 \label{en_dsmd_packet}
 \vspace{-0.5cm}
 \end{minipage}
 
 \end{figure}
 \begin{figure*}[htb!]

 \begin{minipage}{5.6cm}
 \centering
 \epsfig{width=5.6cm,figure=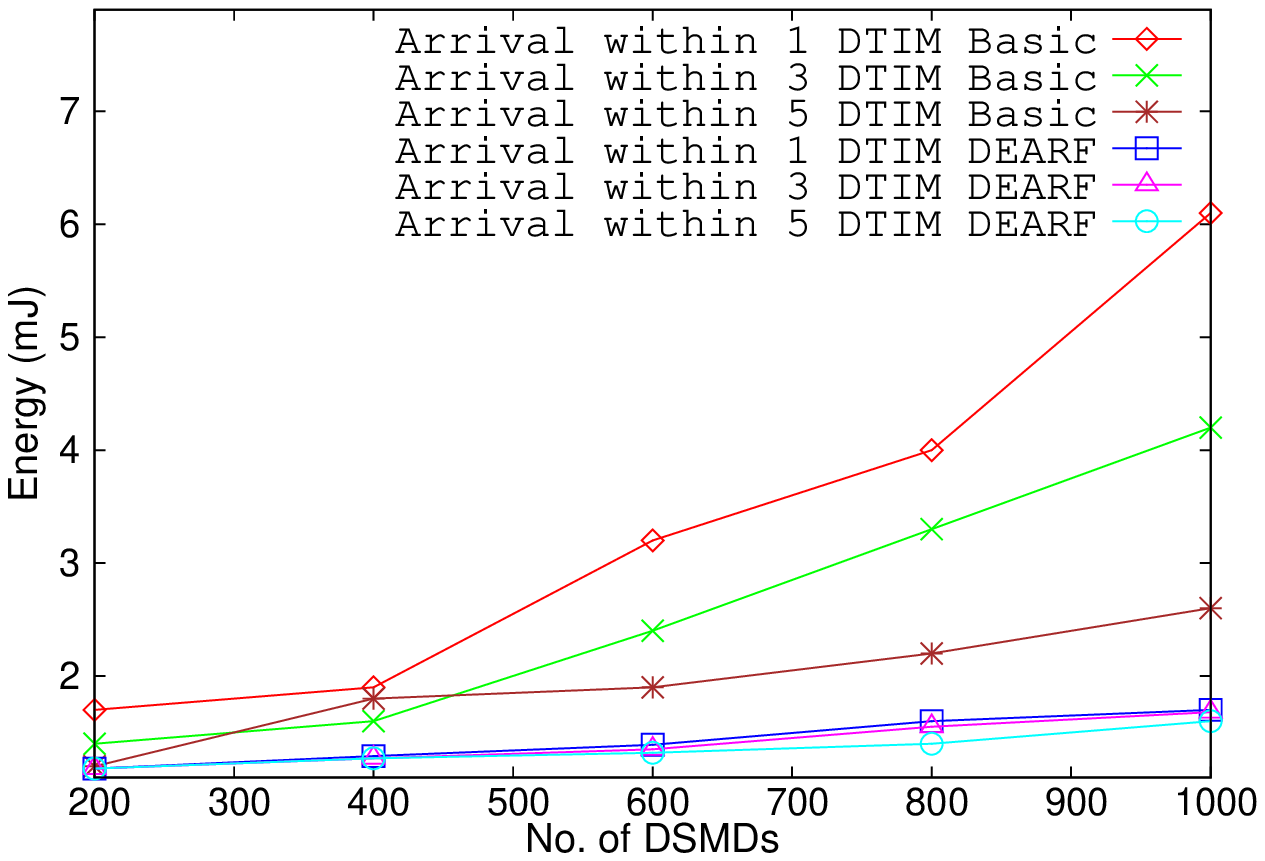}
 \caption{Average energy consumed per DSMD.}
 \label{en_dsmd_total}
 \end{minipage}
 \hspace{0.5cm}
 \begin{minipage}{5.6cm}
 \centering
 \epsfig{width=5.6cm,figure=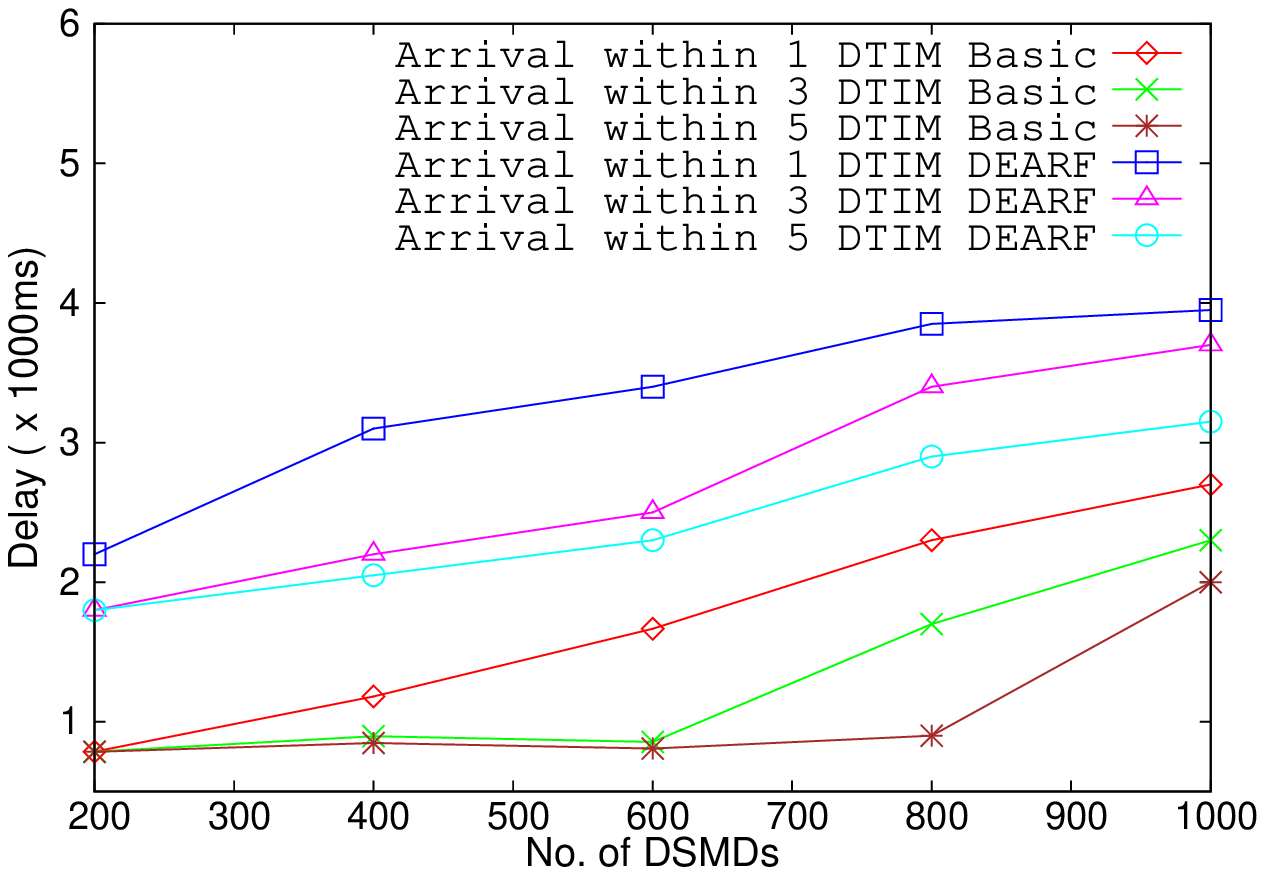}
  \caption{Average delay per packet for DSMDs.}
 \label{delay_dsmd}
 \end{minipage}
 \hspace{0.5cm}
\begin{minipage}{5.6cm}
 \centering
 \epsfig{width=5.6cm,figure=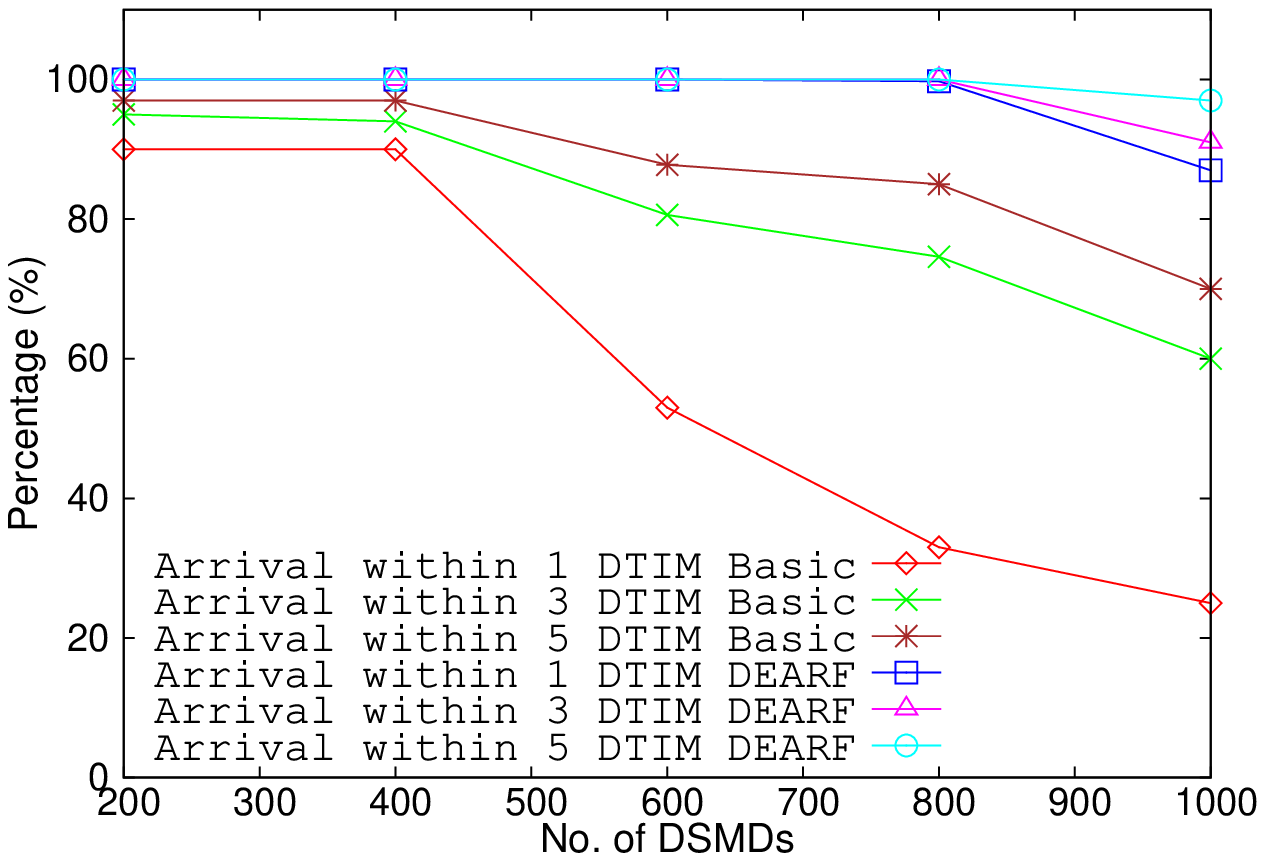}
 \caption{\% of packets transmitted within deadline.}
 \label{percentage_transmitted}
 \end{minipage}
 \vspace{-0.5cm}
 \end{figure*}

 \begin{figure*}[htb!]
 \begin{minipage}{5.6cm}
 \centering
 \epsfig{width=5.6cm,figure=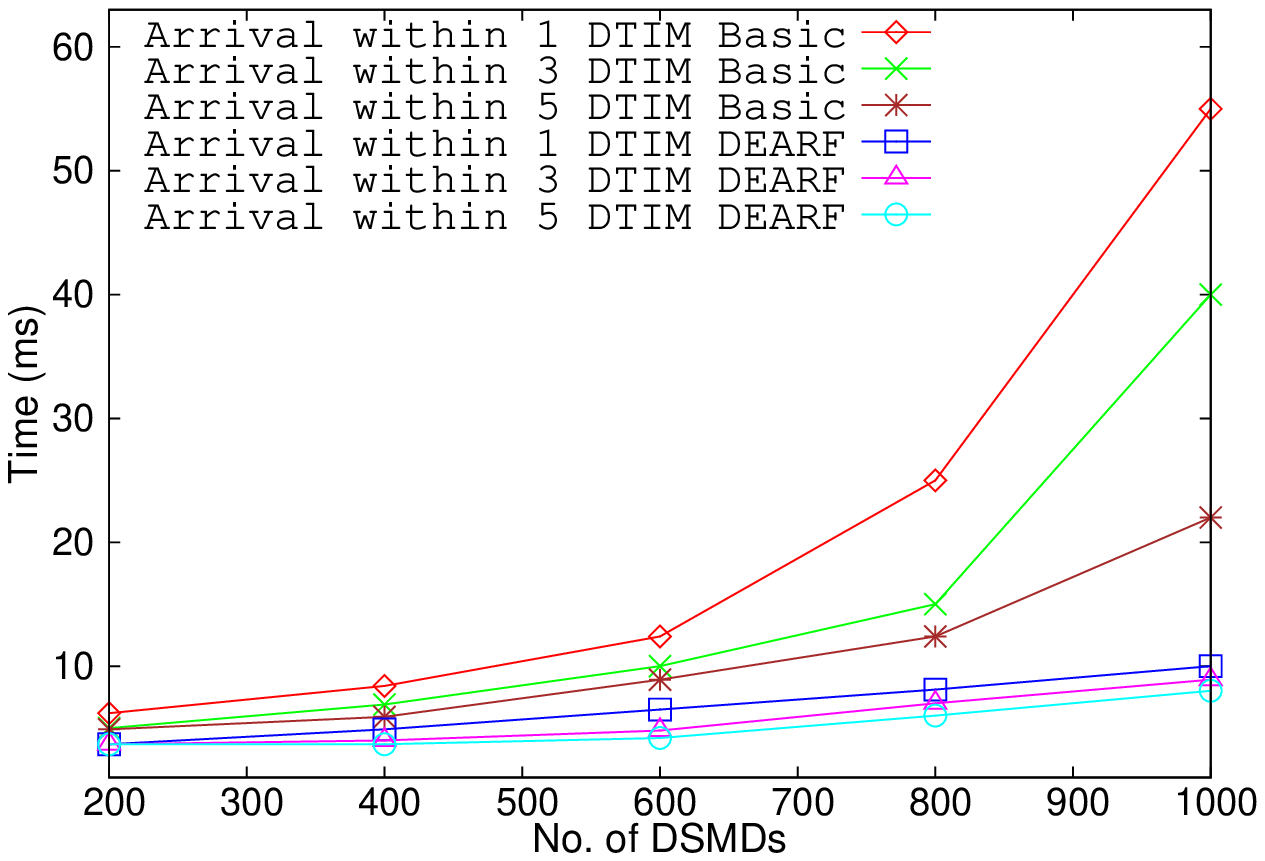}
 \caption{Amount of time a DSMD is active between packet arrival and transmission.}
 \label{active_time}
 \end{minipage}
 \hspace{0.5cm}
 \begin{minipage}{5.6cm}
 \centering
 \epsfig{width=5.6cm,figure=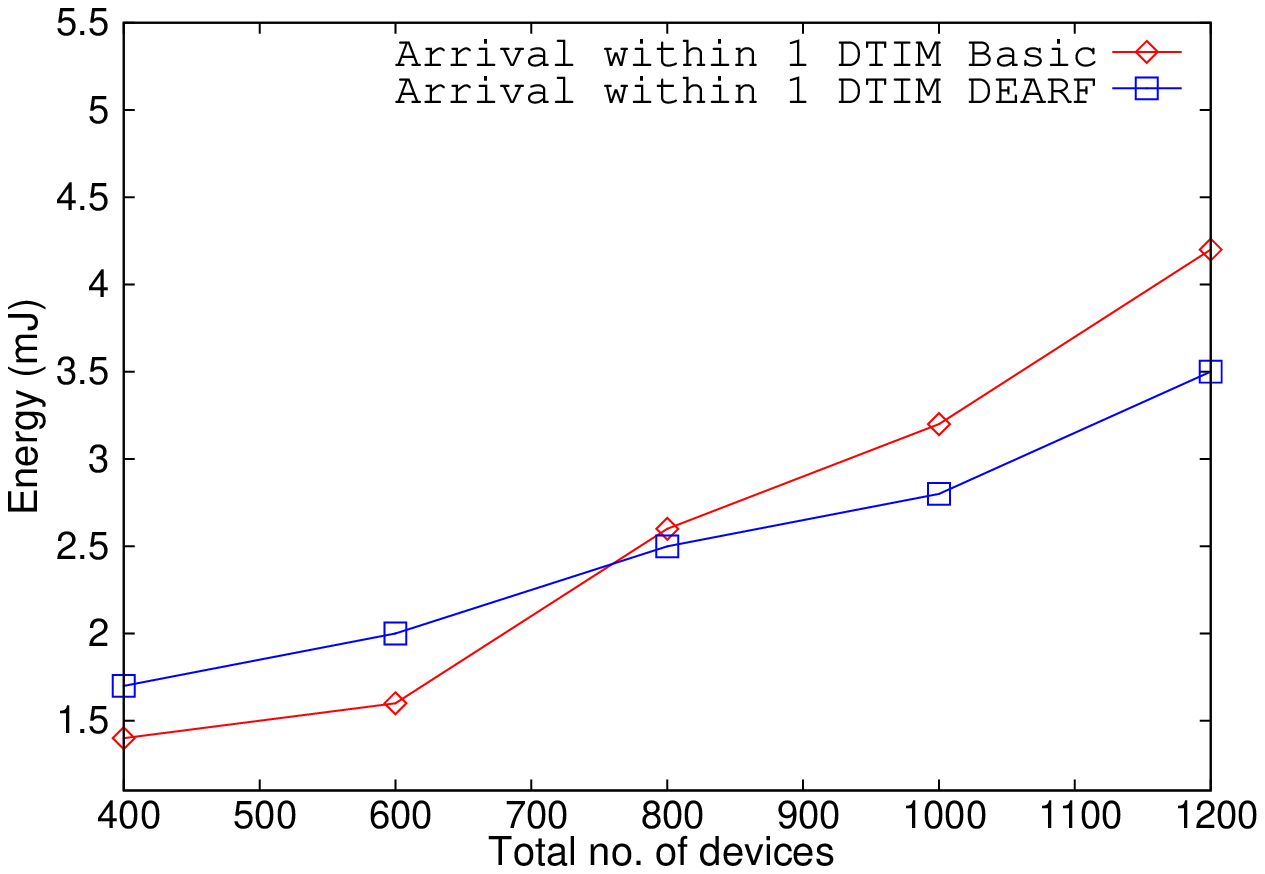}
 \caption{Average energy consumed per packet for non-DSMDs.}
 \label{energy_ndsmd}
 \end{minipage}
 \hspace{0.5cm}
 \begin{minipage}{5.6cm}
 
 \centering
 \epsfig{width=5.6cm,figure=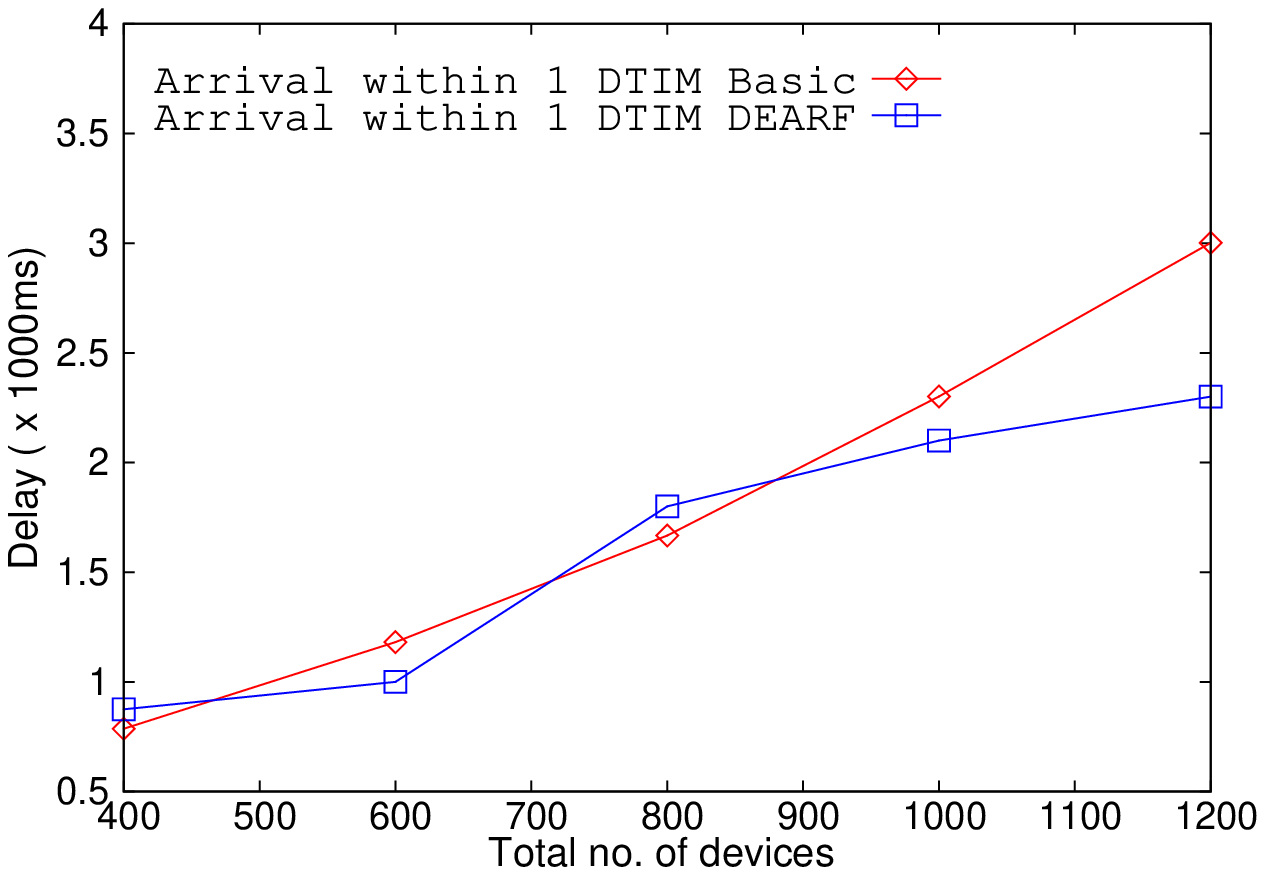}
 \caption{ Average delay per successful packet for non-DSMDs.}

 \label{delay_ndsmd}
 \end{minipage}
 \vspace{-0.5cm}
 \end{figure*}

Figure~\ref{en_dsmd_packet} compares the average amount of energy consumed for sending single packet by a DSMD and Figure~\ref{en_dsmd_total} shows the average energy spent by a DSMD in both schemes. For less load, the Basic scheme has low energy consumption over DEARF scheme. Reasons being, devices in DEARF scheme have to send and receive extra control packets and devices using the Basic scheme easily send their packets due to less contention. As the number of DSMDs crosses 400, we see that our scheme performs better in both cases. Because, as contention in the medium increases, DSMDs face a lot of collisions in Basic scheme, which leads to more retransmissions and medium sensing, hence higher energy consumption. Whereas with our scheme, DSMDs listen to only extra control packets. The DEARF scheme saves between 10-30\% of energy per successful packet, whereas energy savings per DSMD devices is between 100-330\%.

Figure~\ref{delay_dsmd} shows the delay between the arrival of a packet and its successful transmission by a device. The delay for DEARF scheme is found to be in multiples of seconds~(DTIM intervals). This is mainly because, once DSMDs announce they have a packet to send in a particular DTIM, they get a transmission slot only in upcoming DTIM intervals. As load increases they may get a transmission slot in a later DTIM. This leads to increase in average per packet delay for DSMDs in DEARF scheme.

Figure~\ref{percentage_transmitted} shows the percentage of packets transmitted within the requested deadline~(packet delivery ratio). In this figure, we observe that our scheme gives 100\% delivery at low loads, whereas the Basic scheme manages a maximum of 95\%. As the number of devices increases, the DSMDs following Basic scheme drop more number of packets because of excessive retransmissions. The DEARF scheme performs better by giving 90\% deliveries at high load and spontaneous arrival whereas, Basic scheme provides only 27\% deliveries.

Figure~\ref{active_time} shows the average time a DSMD is active~(listening, receiving or transmitting) in between getting a packet from the higher layers to making a successful transmission. As DSMDs increase our scheme gives steady performance as it does not spend time contending for the channel. Whereas, time to stay active for DSMDs performing by Basic scheme increases exponentially, as contention increases these devices suffer increased idle listening and retransmissions. DEARF scheme gives higher average delay, but it does not hinder our motive of reducing energy consumption and reliable packet delivery within deadline.

In Figures~\ref{energy_ndsmd} and~\ref{delay_ndsmd}, we see energy consumption and delay of non-DSMDs. In terms of energy spent per packet, we observe that Basic and DEARF schemes perform closely. DEARF scheme loses out on 10\% savings at low load because of overhead by listening to extra control packets. As the load increases, DEARF scheme helps non-DSMDs to save upto 16\% energy. This is because contention for non-DSMDs increases as number of DSMDs increase in Basic scheme. Whereas, in DEARF scheme DSMDs are given separate access through DRA RAW. In terms of delay both the schemes perform closely at low loads. As load on the system increases, NRA RAW is given less fraction of resources per DTIM which leads to increased waiting times. On the other hand, increase in load leads to increase in contentions in Basic scheme. This leads to quadratic increase in waiting times for packets. The DEARF scheme ends up giving an improvement of 23\% at high loads.

\vspace*{-0.5cm}
\section{Conclusions}
\label{conclusion}
In this paper, we proposed a new hybrid~(DEARF) scheme for supporting delay sensitive devices. In order to achieve this, we split the medium access in two parts viz., contention free access for DSMDs and contention based access for serving non-DSMDs. The delay suffered by devices in proposed DEARF scheme is higher than the devices in Basic scheme, on account of scheduling. However, DSMDs in DEARF scheme could save upto 30\% energy per packet and upto 330\% energy at high loads. The packet delivery ratio improved significantly by achieving 90\% delivery in DEARF scheme over 27\% delivery in Basic scheme at high loads. At the same time, non-DSMDs performing by DEARF scheme saved upto 16\% on energy spent per packet and face 23\% lesser packet delay as compared to the Basic scheme. We conclude that although the higher delay per packet the proposed scheme is able to support DSMDs by giving more reliable delivery and also improving energy savings for both DSMDs and non-DSMDs. Future works include optimizing the resource allocation algorithm using predictive techniques and using intutive grouping techniques to improve CI and DII RAW efficiency.


\



%


\end{document}